\documentclass[opre,nonblindrev]{informs3} 

\OneAndAHalfSpacedXI 

\usepackage{endnotes}
\let\footnote=\endnote
\let\enotesize=\normalsize
\def\notesname{Endnotes}%
\def\makeenmark{$^{\theenmark}$}
\def\enoteformat{\rightskip0pt\leftskip0pt\parindent=1.75em
  \leavevmode\llap{\theenmark.\enskip}}

\usepackage{natbib}
 \bibpunct[, ]{(}{)}{,}{a}{}{,}%
 \def\bibfont{\small}%

 \usepackage[colorlinks=true,breaklinks=true,bookmarks=true,urlcolor=blue,
     citecolor=blue,linkcolor=blue,bookmarksopen=false,draft=false]{hyperref}

\TheoremsNumberedThrough     
\ECRepeatTheorems

\EquationsNumberedThrough    

\usepackage{setspace}
\usepackage{graphicx}
\usepackage{fancyhdr} 
\usepackage{tikz}
\usepackage{todonotes}
\usepackage{setspace}
\usepackage{bbm}
\usepackage{nth}
\usepackage{dsfont,amsfonts}
\usepackage{subfig}
\usepackage{booktabs}
\usepackage{enumerate, enumitem}
\usepackage{cleveref}

\newcommand{\N}{{\mathbb{N}}}
\newcommand{\E}{{\mathbb{E}}}

\renewcommand{\P}{{\mathbb{P}}}

\newcommand{\X}{{\mathcal{X}}}

\newcommand{\z}{{\mathbf{z}}}

\newcommand{\bA}{{\boldsymbol{A}}}

\newcommand{\R}{{\mathds{R}}}

\newcommand{\mF}{{\mathcal{F}}}

\renewcommand{\S}{{\mathfrak{S}}}
\newcommand{\bS}{{\boldsymbol{\mathfrak{S}}}}

\newcommand{\bsS}{{\boldsymbol{S}}}

\newcommand{\VaR}{{\text{VaR}}}
\newcommand{\var}{{\text{var}}}

\newcommand{\ep}{{\varepsilon}}

\NewDocumentCommand{\lux}{mO{\phi}}{%
   \Vert #1 \Vert_{#2} %
}

\begin{document}


\RUNAUTHOR{Ince et al.}

\RUNTITLE{Constructing elicitable risk measures}

\TITLE{Constructing elicitable risk measures}

\ARTICLEAUTHORS{%
\AUTHOR{Akif Ince\footnote{...}}
\AFF{School of Computing and Mathematical Sciences, Birkbeck, University of London, Malet Street, Bloomsbury, London WC1E 7HX, United Kingdom,\\
Standard...
\EMAIL{aince02@student.bbk.ac.uk}} 
\AUTHOR{Marlon Moresco\footnote{corresponding author.}}
\AFF{Escola de Administra\c{c}$\tilde{a}$o, Universidade Federal do Rio Grande do Sul, Porto Alegre, Rio Grande do Sul 90010-460, Brazil, \EMAIL{marlon.moresco@ufrgs.br}}
\AUTHOR{Ilaria Peri}
\AFF{Business School, Birkbeck, University of London, Malet Street, Bloomsbury, London WC1E 7HX, United Kingdom, \EMAIL{i.peri@bbk.ac.uk}}
\AUTHOR{Silvana M. Pesenti}
\AFF{Department of Statistical Sciences, University of Toronto, Toronto, Ontario M5S 3E6, Canada, \EMAIL{silvana.pesenti@utoronto.ca}
\\[3em]
February 25, 2025
}
} 

\ABSTRACT{%
We provide a constructive way of defining new elicitable risk measures that are characterised by a multiplicative scoring function. We show that depending on the choice of the scoring function's components, the resulting risk measure possesses properties such as monotonicity, translation invariance, convexity, and positive homogeneity. Our framework encompasses the majority of well-known elicitable risk measures including all elicitable convex and coherent risk measures. Our setting moreover allows to construct novel elicitable risk measures that are, for example, convex but not coherent. Furthermore, we discuss how higher-order elicitability, such as jointly eliciting the mean and variance or different quantile levels, fall within our setting. 
}%



\maketitle

\section{Introduction}
Functionals (or risk measures) that map random variables or distributions to subset of real numbers are ubiquitous in various fields, including statistic, economics, finance, risk management, and machine learning. For law-invariant functionals, which are those who only depend on the distribution of the random variable, the question on whether they are elicitable has been of considerable interest in the statistical literature; indicatively see \cite{Lambert2008ACM,Gneiting2011,Steinwart2014proceeding} and \cite{Nolde2017AAS}. Elicitability refers to the fact that elicitable functionals admit a representation as the argmin of an expected loss function, the so--called scoring function. Classical examples are the expected value, which is the minimiser of the squared loss, and quantiles, that are the argmin of the expected pinpall loss. Elicitability is of importance for regression analysis, model comparison, and model prediction, as it incentivises truthful prediction, see e.g., \cite{Gneiting2011} and reference therein. From a risk management point of view, elicitability allows for comparable backtests, which are highly important in finance \citep{Acerbi2014Risk,kou2016OR,Nolde2017AAS}. Recently, sensitivity measures tailored to elicitable risk measures have been introduced in \cite{fissler2023EJOR}, elicitability has also played a key role probabilistic opinion pooling \citep{neyman2023OR}, and in risk-aware dynamic decision making \citep{pesenti2024OR}. We also refer the reader to \cite{smith2022OR}, who establishes an intimate connection between the generalised entropy of scoring functions and convex risk measures. 

While a core part of the literature on elicitability focuses on which known functional are elicitable, we provide a different approach in that we propose ways of constructing new elicitable risk measures, an avenue that has not been explored before. Related is \cite{meng2023scores}, who propose novel scoring functions for eliciting multivariate distribution functions, while here we construct new elicitable functionals, which is distinct from eliciting the entire underlying distribution. Specifically, our class of elicitable functionals are characterised by a multiplicative structure of the scoring function. Scoring functions are non-negative functions that penalise the difference between predictions and realisations, with smaller values implying a better model-fit. The first component of our scoring function is an increasing convex function that captures the absolute deviation between realisation and predictions. Thus, the larger the absolute difference between the realisation and prediction, the larger the penalty. The second multiplicative function penalises over and under-estimation asymmetrically. The proposed elicitable risk measures cover well-known elicitable functionals, including the mean, quantiles, expectiles, shortfall risk measures, and generalised quantiles. Moreover, they span the class of elicitable convex and coherent risk measures. 

We prove that the proposed elicitable functionals are well-defined and show necessary (and some sufficient) conditions on the multiplicative scoring function so that the functionals possess properties including monotonicity, translation invariance, positive homogeneity, and convexity. Thus, our framework provides a recipe to construct novel elicitable functionals that possess desirable properties, e.g., are convex but not coherent. We further extend our framework to higher-order elicitability pioneered in \cite{Lambert2008ACM}. While many functionals, such as the variance, are not elicitable on their own, higher-order elicitability is the concept that pairs, or a vector of functionals, is jointly elicitable, i.e. can be written as the argmin over the expected value of a suitable scoring function. We show that jointly eliciting different quantile levels and the (mean, variance) pair fall within our setting. Our framework, which is the first of its kind to explicitly construct old and new elicitable risk measures with specific properties, opens many doors of applications in finance, economics, and risk management as well as in statistics and machine learning. 

The reminder of the paper is structured as follows. \Cref{sec:elicitability} recalls the notion of elicitability, while \Cref{sec:constructing-elicitability} introduces our framework to construct elicitable risk measures using a multiplicative scoring function. In this section we further discuss first-order conditions, show necessary condition for the risk measures to possess different properties, and provide a generalised quantile representation. \Cref{sec:examples} discusses known and new examples, and \Cref{sec:extensions} is devoted to two sets of extensions, the first pertaining to transformation of random variables and Osband's principle, and the second to higher-order elicitability. We conclude in \Cref{sec:conclusion}.

\section{Elicitable functionals}\label{sec:elicitability}
\subsection{Preliminaries}
 Consider an atomless probability space $(\Omega,\mF,\P)$ and let $L^0:=L^0(\Omega, \mF, \P) $ be the space of (equivalent classes of) all random variables, and $L^\infty$ the space of all essentially bounded random variables. We use the notation $\X$ to denote a Banach space satisfying $L^\infty \subseteq \X \subseteq L^0$. For a random variable $X \in L^0$, we denote its cumulative distribution function (cdf) by $F_X(x) := \P(X\leq x)$, $x \in \R$, and its (left-continuous) quantile function by $F^{-1}_X (\alpha) := \inf\{ x \in \R :  F_X (x) \ge \alpha\}$, $\alpha \in (0,1)$.

Key to the exposition are Orlicz spaces, thus we next recall the definition of a Young function. 

\begin{definition}[Young function]
A function $\phi : [0, \infty) \to [0, \infty]$ is a Young function, if it is left-continuous, convex, and satisfies $\lim_{x \to 0} \phi(x) = \phi(0) = 0$ and $\lim_{x \to \infty} \phi (x) = \infty$.     
\end{definition}
By definition, a Young function is non-decreasing and continuous, except potentially at a single point where it jumps to $\infty$. Given a Young function $\phi$, its associated Orlicz space is defined by
\begin{equation*}
    L^\phi := \big\{ X \in L^0 ~ \big|~  \E\big[\phi \big(c |X| \big) \big] < \infty, \text{ for some } \; c >0\big\}\,,
\end{equation*}
and its Orlicz Heart --- a subset of the Orlicz space --- is
\begin{equation*}
    H^\phi := \big\{ X \in L^0 ~ \big|~  \E\big[\phi \big(c |X| \big) \big] < \infty, \text{ for all } c >0\big\}\,.
\end{equation*}
We refer to \cite{harjulehto2019generalized} for a systematic treatment of Orlicz spaces and hearts.

\subsection{Elicitability}
Let $R\colon \X \to A$, $A \subseteq \R$, be a law-invariant functional mapping random variables to subsets of the real line, that is $R$ migth be set-valued. Recall that a functional $R$ is law-invariant, if $R[X] = R[Y]$, whenever $F_X(\cdot) = F_Y(\cdot)$. Of interest are elicitable law-invariant functionals, recalled next, and we refer to \cite{Gneiting2011} and \cite{Ziegel2016} and reference therein.

\begin{definition}[Consistency and Elicitability]
\label{def:elicitable}
For an action domain $A \subseteq \R$, a scoring function (or simply score) is a measurable map $S:A\times \R \to [0,\infty]$. Given a law-invariant, potentially set-valued functional, $R: \X \to A $, a score is called 
\begin{enumerate}[label = \roman*)]
    \item consistent for $R$, if for all $X \in \X$ and for all $z \in A $ it holds that
    \begin{equation}\label{eq:consistency}
        \E\big[ S\big(R[X],X\big) \big]  \leq  \E \big[ S(z,X) \big]\, .
    \end{equation}

    \item strictly consistent for $R$, if equality in \eqref{eq:consistency} holds only if $z \in R[X] $.
            
\end{enumerate}
We say that $R$ is elicitable if there exists a consistent scoring function for $R$, in which case $R$ admits the representation
\begin{equation*}
    R[X]  = \argmin_{z \in A} \E \big[ S(z,X) \big] \quad \text{for all} \quad X \in \X\,. 
\end{equation*}
\end{definition}

As we introduce new elicitable functionals, of interest is which properties these new functionals satisfy. Thus, we recall desirable properties of risk measures / functionals using the convention that positive values correspond to losses. We refer the interested reader to  \cite{ArtznerDelbaenETAL1999} and \cite{Frittelli2002JBF} for discussions and interpretaion of these properties. Note that while $R$, the elicitable functionals may be set-valued, risk measures map to the real line.

\begin{definition}\label{def-rm-properties}
    A functional $r : \X \to \R $ may satisfy some of the following properties:
    \begin{enumerate}[label = \roman*)]
        \item monotonicity: if $X \leq Y$ $\P$-a.s. implies that $r[X] \leq r [Y]$ for all $X,Y \in \X$.
        \item translation invariance: if $r[X+c]=r[X] + c$ for all $c \in \R$ and  all $X \in \X$.
        \item positive homogeneity: if $r[\lambda X] = \lambda r[X]$ for all $ \lambda \in [0,\infty)$, and all $X \in \X$.
        \item convexity: if $r[\lambda X + (1-\lambda) Y] \leq \lambda r [ X] + (1-\lambda) r[Y]$ for all $\lambda \in [0,1]$, and all $X,Y \in \X$.
        \item star-shaped: if $r[\lambda X] \le \lambda r[X]$ for all $X \in \X$ and $\lambda \in [0,1]$.
    \end{enumerate}        
\end{definition}

\section{Constructing elicitable functionals}\label{sec:constructing-elicitability}

\subsection{Elicitable functional with multiplicative scores}
This section introduces families of new elicitable functionals, for which we need additional notation. We say a function $f: \R^2 \to \R$ is \textit{convex} if it is jointly convex, i.e., the epigraph $\{(z_1, z_2, r) \in \R^3 \;|\; r \ge f(z_1, z_2)\}$ is a convex set. Throughout, we simply say that $f$ is convex, meaning jointly convex, and explicitly state if we only require convexity in one of the components. We use the notations $\partial_x f(z, x):= \frac{\partial}{\partial x}f(z,x) $ and $\partial_z f(z, x):= \frac{\partial}{\partial z} f(z,x)$ to denote partial derivatives for a function $f\colon \R^2 \to \R$, whenever they exist. We denote the left and right derivative by $\partial_x^- f(z, x):= \frac{\partial^-}{\partial x}f(z,x) $, respectively, by $\partial_x^+ f(z, x):= \frac{\partial^+}{\partial x}f(z,x) $. For a univariate function $g \colon \R \to \R$ we simply write $g'(z):= \frac{d}{d z} g(z)$, if the derivative exists, and $g'(|z-x|)$ is understood as the derivative evaluated at $|z-x|$, i.e. $g'(|z-x|):= g'(y)\big|_{y = |z-x|}$.

Important to the novel elicitable functionals is the notion of accuracy rewarding, see e.g.,  \cite{Lambert2008ACM} and \cite{Pesenti2024ORL} for interpretation and its connection to risk measures.

\begin{definition}[Accuracy rewarding]
We call a function $f \colon \R^2 \to \R$ \textit{accuracy rewarding} if it satisfies $(i)$ $f(z,x)=0$ if and only if $z=x$, $(ii)$ the function $z \mapsto f(z,x)$ is non-decreasing in $z$ whenever $z>x$, and $(iii)$ and the function $z \mapsto f(z,x)$ is non-increasing in $z$ whenever $z<x$.     
\end{definition}

Equivalently, $f$ is accuracy rewarding, if $(i)$ holds and if $f$ is non-decreasing in $x$, for $z<x$, and non-increasing in $x$, for $z>x$. Next, we introduce our new class of elicitable functionals.

\begin{definition}[Elicitable functional with multiplicative scores]\label{definition of main functional}
    Let $\phi$ be an increasing Young function satisfying $\phi(x) = 0$ if and only if $x =0$, and let $\mu \colon \R^2 \to [0,1]$ be an accuracy rewarding convex function. Then for all $X \in H^\phi$, we define the functional
    \begin{equation}\label{eq:rm-def}
        \rho[X] := \argmin_{z \in \R}\,  \E \big[\phi(|z-X|) \mu(z,X) \big]\,. 
    \end{equation}
\end{definition}
If $\rho[X]$ is set-valued, one can take either the left or the right endpoint to obtain a risk measure, i.e. $\rho^-[X] := \inf \rho[X]$ or $\rho^+[X] := \sup \rho[X]$. 
Clearly, the functional $\rho$ is elicitable with score $\S \colon \R^2 \to [0, \infty)$, given by
\begin{equation}\label{eq:score}
   \S(z, x) := \phi(|x - z|)\mu(z, x),
\end{equation}
The choices of $\phi$ and $\mu$ in \eqref{eq:score} do not admit a unique representation, as one can multiply $\mu$ and divide $\phi$ by a positive number strictly smaller than one. In many situations, however, the functionals $\phi$ and $\mu$ arise naturally. 
\begin{lemma}
    The score $\S$ is accuracy rewarding and consistent for $\rho$.
\end{lemma}
\proof{Proof.} 
   First we show accuracy rewarding. From the definition of $\phi$ and $\mu$, it follows directly that \( \S(z, x) = 0 \) if and only if \( x = z \). Next, consider the case where \( z > x \). In this scenario, both \( |x - z| \) and \( \mu(z, x) \) are non-decreasing. Given that \( \phi \) and \( \mu \) are non-negative, and \( \phi \) is non-decreasing, it follows that \( \S(z, x) \) is also non-decreasing. A similar reasoning holds when \( z < x \).
   Consistency follows by definition of $\rho$.
   \hfill \Halmos
\endproof

The first set of results pertain to the existence and finiteness of the new functional.
\begin{proposition}\label{prop: existence}   
Let $X \in H^\phi$ and assume that $ \S(z, x)$  is lower-semicontinuous in $z$, then
\begin{enumerate}[label = \roman*)]

    \item\label{prop: existence finite} $\rho[X]$ is non-empty and compact. Hence, $-\infty < \rho^-[X] < \rho^+[X] <\infty$ for all $X \in H^\phi$.

    \item\label{prop: existence bounds} $\rho[X] \subseteq [\inf X, \sup X]\cap \R$.

    \item\label{prop: existence interval} If $\S$ is convex in $z$, then $\rho[X]$ is a closed interval,

    \item \label{prop: zero} it holds that $\rho[0] = 0$.
\end{enumerate}
\end{proposition}

\proof{Proof.} 
    \Cref{prop: existence finite}, consider the function $f_X(z) := \E[\phi(|X-z| )\mu(z,X)]$, as $X-z \in H^\phi$ for any $X \in H^\phi$, it follows that $\E[\phi(|X-z| )]$ is finite.  Further, as $\phi(|X-z| ) \geq 0$, it holds that $\phi(|X-z| )\mu(z,X) \leq \phi(|X-z| )$ $\P$-a.s. Moreover, $\phi$ is level bounded by assumption, and, as $\mu$ is accuracy rewarding, it holds that $\lim_{z \to \infty}\mu(z,x) >0$ for all $x \in \R$.  Hence, $\lim_{y \to \infty} \phi(|X-z| )\mu(z,X) = \infty $, i.e. $f_X(z)$ is also level bounded. Therefore, each sublevel set is bounded and closed, hence, compact. Thus, by the extreme value theorem of semicontinuous functions,  $f_X(z)  $ attains its minimum, and the set $\argmin_{z \in \R}f_X(z)  = \{z \in \R : f_X(z) \leq \min_{y\in \R} f_X(y)\} $ is non-empty and compact. 
        
    \Cref{prop: existence bounds} follows directly from the fact that $\S$ is accuracy rewarding.

    \Cref{prop: existence interval}  follows from convexity of the sublevel sets.   

    \Cref{prop: zero} note that $\rho[0] = \argmin_{z \in \R}\,  \E \big[\phi(|z-0|) \mu(z,0) \big]$. Now since $\phi(z) = 0$ if and only if $z = 0$, the argmin is attained at the value 0.
    \hfill \Halmos
\endproof

As elicitable functionals are defined as the minimiser of a convex optimisation problem, they are often found by solving the corresponding first-order condition (f.o.c.).
\begin{proposition}[First-order condition]\label{prop: basic properties}
Let $X \in H^\phi$ and assume that $ \S(z, x)$  is convex in $z$, then the following hold:
\begin{enumerate}[label = \roman*)]
    \item\label{prop: basic properties singleton} 
    If $z \mapsto \E[\S(z,X)]$ is differentiable with strictly increasing derivative, then $\rho[X]$ is a singleton.

    \item\label{prop: basic properties first order condition} 
    We have $z^* \in \rho[X]$ if and only if $z^*$ satisfies the following f.o.c.
\begin{align*}
     \E \big[  (  1_{X < z^*}  - 1_{X\geq z^*} )& \mu(z,x) \partial^-_z \phi(|X-z^*|)  + \phi(|X-z^*|)\partial^-_y \mu(z^*,X) \big] 
     \\
      &
      \leq 0 
     \leq
     \E \big[ (  1_{X \leq z^*}- 1_{X> z^*}) \mu(z,x) \partial^+_z \phi(|X-z^*|)  + \phi(|X-z^*|)\partial^+_y \mu(z^*,X) \big]\,  .
\end{align*}
If further the derivative of $\phi$  exists, i.e. $\partial^-_z \phi (\cdot)= \partial^+_z \phi (\cdot)$, the f.o.c. becomes 
\begin{align}\label{eq:foc-2}
      \E  \big[   \phi(|X-z^*|){\partial^-_z \mu(z^*,X)}        \big]  
      &\leq 
         \E \big[ (  1_{X\geq z^*}-1_{X < z^*}) \mu(z^*,X)\phi'(|X-z^*|)         \big] 
         \\ &\leq 
       \E  \big[\phi(|X-z^*|)\partial^+_z \mu(z^*,X)  \big]\,. 
       \notag
\end{align}    
If additionally, the derivative of $\mu$ with respect to $z$ exists, then the f.o.c. reduces to 
\begin{equation*}
      2F_X(z^*)\, \E\big[   \phi'(|X-z^*|) \mu(z^*,X)    ~\vert ~X \leq z^*     \big]       
      = \E \big[  \phi'(|X-z^*|) \mu(z^*,X)- \phi(|X-z^*|) \partial_z \mu(z^*,X)        \big]\,.
\end{equation*}
\end{enumerate}
\end{proposition}

\proof{Proof.}
We first prove \Cref{prop: basic properties first order condition}. Since $f_X$ is convex, it holds that $z^* \in \rho[X]$, if and only if
    \begin{equation*}
        0 \in \big[ \, \partial^-_z f_X(z^*),\partial^+_z f_X(z^*)  \,  \big]\,.
    \end{equation*}
Dominated convergence yields 
\begin{align*}
         \partial^-_z f_X(z^*) 
         &= \E  \big[  \partial^-_z \big(\phi(|X-z^*|)\mu(z^*,X)   \big)\big]
         \\ &=
         \E \big[ (  1_{X < z^*} - 1_{X\geq z^*} ) \mu(z^*,X)\partial^-_z\phi(|X-z^*|)  + \phi(|X-z^*|) \partial^-_z \mu(z^*,X)       \big] \,,
\end{align*}
and similarly
\begin{align*}
    \partial^+_z f_X(z^*) 
     &=
     \E \big[ ( 1_{X \leq z^*}- 1_{X> z^*} ) \mu(z^*,X)\partial^+ _z\phi(|X-z^*|)  + \phi(|X-z^*|) \partial^+_z \mu(z^*,X)       \big] \,.
\end{align*}
The second set of inequalities follows from the first ones by noting that $\phi'(z) = \partial^-_z\phi(z) = \partial^+_z\phi(z)$. To see the last set of inequality, since $\partial^-_z\mu(z, x) = \partial^+_z \mu(z, x)$, \eqref{eq:foc-2} is equivalent to 
\begin{align*}
     \E  \big[   \phi(|X-z^*|){\partial_z \mu(z^*,X)}        \big] 
    &=
    \E \big[ (  1_{X\geq z^*}-1_{X < z^*}) \mu(z^*,X)\phi'(|X-z^*|)         \big] 
    \\
    &= 2\, \E \big[   1_{X\geq z^*} \mu(z^*,X)\phi'(|X-z^*|)         \big]
    -\E \big[\mu(z^*,X)\phi'(|X-z^*|)         \big]
    \\
    &= 2\,F_X(z^*) \, \E \big[  \mu(z^*,X)\phi'(|X-z^*|)         ~|~ X \le z^*\big]
    -\E \big[\mu(z^*,X)\phi'(|X-z^*|)         \big]\,.
\end{align*}

\Cref{prop: basic properties singleton} by differentiability of the expected score, it holds $f_X'(z^*) :=  \partial^-_z f_X(z^*)=\partial^+_z f_X(z^*)$, and the f.o.c. reduces to $f_X'(z^*)=0$. Uniqueness follows by the strictly increasingness of the expected score.
\hfill \Halmos
\endproof

When the derivatives of $\phi$ and $\mu$ exist, the f.o.c. simplifies significantly. Throughout, we set
\begin{align}\label{eq: f.o.c. func h}
     h(z,x) := -\partial_z\S(z,  x)
     =
     (  1_{x\geq z}-1_{x < z}) \mu(z,x)\phi'(|x-z|) -  \phi(|x-z|) \partial_z \mu(z,x)\,,
 \end{align}
so that $\rho[X]$ is the set of solutions to $\E[h(\rho[X],X) ] = 0 $.

\subsection{Properties }
In this section we provide necessary and some sufficient requirements on $\phi$ and $\mu$ such that our proposed elicitable functional has the properties given in \Cref{def-rm-properties}.

\begin{theorem}\label{prop: basic properties of functional}
Let $X \in H^\phi$ and $\rho[X]$ be non-empty and bounded. Then the following hold:
\begin{enumerate}[label = \roman*)]
    \item\label{prop: basic properties of functional translation invariance} If $\mu(z,x+c) = \mu(z-c,x)$ for all $x,z,c \in \R$, then  $\rho^+[X]$ and $\rho^-[X]$ are translation invariant.

    \item\label{prop: basic properties of functional monotonicity} If $\S$ is convex, then $\rho^+[X]$ and $\rho^-[X]$ are monotone.
    
    \item\label{prop: basic properties of functional positive homogeneity} If $\mu(\lambda z,\lambda x) = \lambda^p\mu(z,x)$ and $\phi(\lambda y) = \lambda^q \phi (y)$  for all $\lambda,y\ge 0$, for all $x,z \in \R$,  and some $p,q \in \R $,  then $\rho^+[X]$ and $\rho^-[X]$ are positive homogeneous.

    \item\label{prop: basic properties of functional convexity} If  $z \to  \E[\S(z, X)]$ is, for any $ X\in H^\phi$, differentiable with strictly increasing derivative and $h$ is convex, then $\rho[X] = \rho^+ [X]= \rho^-[X]$, and $\rho[\cdot]$ is convex.
    
    \item\label{prop: basic properties of functional star-shaped} If  $z \to \E[\S(z, X)]$ is, for any $ X\in H^\phi$, differentiable with strictly increasing derivative and $h$ is star-shaped, then $\rho[X] = \rho^+[X] = \rho^-[X]$, and $\rho[\cdot]$ is star-shaped.
\end{enumerate}
\end{theorem}

\proof{Proof.} 
    \Cref{prop: basic properties of functional translation invariance}  is straightforward. 

    \Cref{prop: basic properties of functional monotonicity} define $f(z,X) := \partial^-_z \E[\S(z,X)]$ and $g(z,X) := \partial^+_z \E[\S(z,X)]$. Since $\E[\S(z,X)]$ is convex in $z$, the functions $f$ and $g$ are (for all $\omega\in \Omega$) non-decreasing in $z$. This yields 
    \begin{align*}
        \rho^-[X] &= \inf \{z\in \R: g(z,X) \geq 0 \geq f(z,X)\} =  \inf \{z\in \R: g(z,X) \geq 0\} 
        \quad \text{and}
        \\ 
        \rho^+[X] &= \sup \{z\in \R: f(z,X) \leq 0 \leq g(z,X) \} = \sup \{z\in \R: f(z,X) \leq 0\}\,.
        \end{align*}
        Additionally, convexity and accuracy rewarding of $\S$ yield that $f$ and $g$ are non-increasing in $X$, this implies that for $X \leq Y$, the following set inclusions holds $\{z\in \R: g(z,X) \geq 0\} \supseteq \{z\in \R: g(z,Y) \geq 0\}$ and $ \{z\in \R: f(z,X) \leq 0\} \subseteq  \{z\in \R: f(z,Y) \leq 0\}$. Hence, $\rho^-$ and  $\rho^+$ are monotone. 

        \Cref{prop: basic properties of functional positive homogeneity} follows since for $\lambda \ge 0$ it holds
        \begin{align*}
            \rho[\lambda X] &= \argmin_{z \in \R} \E \big[\phi(|z-\lambda X|) \mu(z, \lambda X) \big]
            \\
            &=  \argmin_{z \in \R} \E \big[ \lambda^q \phi \big( \big|\tfrac{z}{\lambda}- X\big|\big) \lambda^p \mu\big(\tfrac{z}{\lambda}, X\big) \big]
            \\ &=
            \argmin_{\lambda z \in \R} \lambda^{q+p}\E \big[\phi(|z- X|) \mu(z,  X) \big]  
            \\ &=
            \lambda \argmin_{ z \in \R} \E \big[\phi(|z- X|) \mu(z,  X) \big]  
            \\ &=  \lambda \rho[ X]\,.
        \end{align*}

    \Cref{prop: basic properties of functional convexity} by \Cref{prop: basic properties} $\rho[X]$ is a singleton and the unique solution to \Cref{eq: f.o.c. func h}, i.e. to $\E [ h(\rho[X],X)] =0$. By convexity of $h$ we have
    \begin{align*}
         \E [ h(\lambda\rho[X]+ (1-\lambda)\rho[Y],\, \lambda X+(1-\lambda)Y)]
         & \le
         \lambda \, \E \big[ h\big(\rho[X],X\big)\big] + (1-\lambda)\, \E\big[h\big(\rho[Y],Y\big)\big] 
         \\
         &= 0
         \\
         &= \E \big[\,  h\big(\rho[\lambda X+ (1-\lambda) Y],\, \lambda X+(1-\lambda)Y\big)\, \big]\,.
    \end{align*}
This yields, 
\begin{align*}
   \E \big[\,  h\big(\rho[\lambda X+ (1-\lambda) Y],\, \lambda X+(1-\lambda)Y\big)\, \big]
   \leq
   \E \big[ h\big(\rho[\lambda X+ (1-\lambda) Y],\lambda X+(1-\lambda)Y\big)\big]\,. 
\end{align*} 
The non-increasing behaviour of $h$ in the first argument, recall that $h(z, x) = - \partial_z \S(z, x)$, gives convexity of $\rho[\cdot]$.

\Cref{prop: basic properties of functional star-shaped}, from \Cref{prop: basic properties}, \ref{prop: zero} we have $\rho[0]=0$. Repeating the arguments of \Cref{prop: basic properties of functional convexity} with $Y=0$ $\P$-a.s. concludes the proof.
\hfill \Halmos
\endproof

Note that if $\mu$ is positive homogeneous, i.e. $\mu(\lambda z, \lambda \mu) = \lambda^p \mu(z,x)$ for all $x,z \in \R$ and $\lambda>0$, then, as $\mu \in [0,1]$, it must holds that $p=0$. That is \Cref{prop: basic properties}, \ref{prop: basic properties of functional positive homogeneity} requires $\mu$ to be scale invariant.

The next result shows requirements on the score $\S$, so that $\rho$ is either positive homogeneous, convex, or star-shaped.

\begin{theorem}
    Let $X \in H^\phi$, $\rho[X]$ be non-empty and bounded, and let $h$ exists. Then the following hold:
    \begin{enumerate}[label=$\roman*)$]
        \item\label{prop only if p.h.} If $\rho$ is positive homogeneous, then the score $\S$ is $p$-homogeneous for some $p \in\R$, i.e. $\S(\lambda z, \lambda x) = \lambda^p \S(z,y)$, for all $\lambda\ge0$, and all $x,z \in \R$. 
        
        \item\label{prop only if conv} If $\rho$ is convex, then $h$ is convex in its second argument.
        \item\label{prop only if star} If $\rho$ is star-shaped, then $h$ is star-shaped, i.e. $h(\lambda z, \lambda x) \leq \lambda h (z,x)$ for all $z,x \in \R$, $\lambda \in [0,1]$.
        
    \end{enumerate}
\end{theorem}

\proof{Proof.}     
 \Cref{prop only if p.h.}, remember that $z \in \rho[X]$, if it is a solution to $\E[h(z,X)]=0$. Take $x<z<y$ and let $X = x1_B + y(1-1_B)$, where the set $B$ satisfies $\P(B)=\frac{h(z,y)}{h(z,y)-h(z,x)}=:q$. Since $\S$ is accuracy rewarding, we have that $q \in [0,1]$. Then, $\E[h(z,X)] = qh(z,x) + (1-q)h(z,y) = 0$. Therefore, $z \in \rho[X]$, and by positive homogeneity, $\lambda z \in \rho[\lambda X]$ for any $\lambda\geq 0$. This yields that $\E[h(\lambda z,\lambda X)] =0=\E[h(z,X)]$, which is equivalent to \begin{align*}
      \frac{h(\lambda z,\lambda y)}{h(z,y)} =\frac{h(\lambda z,\lambda x)}{h(z, x)}=: g(\lambda).
 \end{align*}
 Then, for any $z,x \in \R$, $h(\lambda z, \lambda x) = g(\lambda)h(z,x)   $. Additionally,  as $g$ is a multiplicative Cauchy equation, which by 
 \cite{kannappan2009functional} Theorem 1.49, is positive homogeneous, $h$ is positive homogeneous. It then follows that $\S$ is positive homogeneous.

\Cref{prop only if conv}, by contradiction assume that $h$ is not convex in its second argument. Then there exists $x,x',z \in \R$ such that $ \frac{h(z,x) + h(z,x')}{2}< h\big(z,\frac{x'+x}{2}\big)$. Thus, there exists a $y \in \R$ satisfying 
\begin{align*}
\lambda \, h(z,y) + \tfrac{1-\lambda}{2} \, \big(h(z,x)+h(z,x')\big) <0 < \lambda \, h(z,y) + (1-\lambda) \, h \big( z, \tfrac{x+x'}{2}\big),
\end{align*}
for some $\lambda \in (0,1)$. Next consider the random variables $X, Y$ each taking values $x,x',y$, and satisfying $\P(X=y,Y=y)=\lambda$ and $\P(X=x,Y=x')=\P(X=x',Y=x) = \frac{1-\lambda}{2}$. Then, 
\begin{align*}
    \E[h(z,X)]  &= \E[h(z,Y)] 
    = \lambda h(z,y) + \tfrac{1-\lambda}{2} ( h(z,x)+h(z,x') )
    <0,
\end{align*}
This implies that $\rho^+ [X]=\rho^+ [Y] < z$. Moreover, we have
\begin{align*}
    \E\big[h\big(z, \tfrac{X+Y}{2}\big)\big] 
    &=
      \lambda h(z,y)  +   (1-\lambda) h \big(z, \tfrac{x+x'}{2}\big) 
       >0\,,
\end{align*}
which imply that $\rho^- \big( \frac{X+Y}{2} \big) > z$. Hence, $\rho$ can not be convex. 

\Cref{prop only if star} follows similar steps to \Cref{prop only if conv}, noting that if $h$ is not star-shaped in the second argument, then there exists $x,z \in \R$ such that $ \frac{1}{2}h(z,x)< h(z,\frac{x}{2})$. Using similar arguments as in \Cref{prop only if conv} provides a contradiction.
\hfill \Halmos
\endproof

\subsection{Generalised  quantile representation}
This section provides a representation of the elicitable functionals as an infimum, a representation akin to that of a quantile function. While this representation is well-known for quantiles and Lambda-quantiles, we show that a similar representation holds for all elicitable functionals of the form \eqref{eq:rm-def}.

\begin{theorem}\label{theo: inf rep}
    Let $\S$ be convex. Then it holds for all $X\in H^\phi$, that
    \begin{align*}
        \rho^+[X] = \inf\Big\{z \in \R: \frac{F_X(z)}{1-F_X(z)} > G_X^-(z) \Big\},
        \\ 
\rho^-[X] = \inf\Big\{z \in \R: \frac{F_X(z)}{1-F_X(z)} \ge G_X^+(z)  \Big\},
    \end{align*}
where $G_X^+$ is defined by 
\begin{align*}
   G^+_X(z) := \dfrac{ \E \big[  \mu(z,X)  \partial^+_z\phi( X-z) - \partial^+_z \mu(z,X) \phi(X-z) ~\big\vert~ X\geq z   \big] }{ \E \big[  \mu(z,X)  \partial^+_z\phi( z-X)+ \partial^+_z \mu(z,X) \phi(z-X) ~\big\vert~ X < z  \big]}\, .
\end{align*}
$G_X^-$ is defined analogously to $G_X^+$, with right derivatives ($ \partial^+_z$) replaced by left derivatives ($ \partial^-_z)$. 
Moreover, the functions $G_X^+$ and $G_X^-$ satisfy  $(1-F_X (z))G_X^-(z) \in [0,1)$ and $(1-F_X(z))G^+_X(z) \in (0,1]$, for all $z \in \R$.
\end{theorem}

\proof{Proof.} 
By \Cref{prop: basic properties}, $\rho[X]$ is a closed interval for all $X \in H^\phi$. Using the same notation as in the proof of \Cref{prop: basic properties of functional}, \ref{prop: basic properties of functional monotonicity}, we can write $g(\cdot, \cdot)$ as 
\begin{align*}
    g(z,X) &= 
\E \Big[ \mu(z,X) \Big(  1_{X < z}  \partial^+_z\phi( |X-z|) - 1_{X\geq z}\partial^+_z\phi( |X-z|) \Big)
\\
&  \quad 
+  \partial^+_z \mu(z,X) (\phi(|X-z|) 1_{X < z} + \phi(|X-z|) 1_{X \geq z} )      \Big]
\\
&=
\big( 1-F_X(z) \big)\,  \E \big[ \partial^+_z \mu(z,X) \phi(X-z)  - \mu(z,X)  \partial^+_z\phi( X-z)  ~\big\vert~ X\geq z \big]
\\ 
&\quad+ 
 F_X(z) \, \E \big[  \partial^+_z \mu(z,X) \phi(z-X) + \mu(z,X)  \partial^+_z\phi( z-X) ~\big\vert~ X< z \big]
\end{align*}
Now note that $\mu(z,X),  \partial^+_z\phi$, and $\phi$ are non-negative and that $ \partial^+_z \mu(z,X) $ is non-negative on $X\leq z$.
Thus, $g(z,X)\geq 0$ is equivalent to
\begin{align*}
    \frac{F_X(z)}{1-F_X(z)} &\geq \dfrac{ \E \big[  -\partial^+_z \mu(z,X) \phi(X-z) + \mu(z,X)  \partial^+_z\phi( X-z)  \big\vert X\geq z \big] }{ \E \big[  \partial^+_z \mu(z,X) \phi(z-X) + \mu(z,X)  \partial^+_z\phi( z-X) \big\vert X< z \big]}
    = G^+_X(z).
\end{align*}
This implies that $\rho^-[X] = \inf\big\{z \in \R:\frac{F_X(z)}{1-F_X(z)} \ge G^+_X(z)  \big\}$. Similarly, by noting that $\rho^+[X] = \sup \{z \in \R: f(z,X) \leq 0 \} = \inf \{ z \in \R: f(z,X) > 0\}$ we have that $\rho^+[X] = \inf\big\{z \in \R: \frac{F_X(z)}{1-F_X(z)}> G^-_X(z) \big\}$.

Recall from \Cref{prop: basic properties} that $-\infty < \rho^-[X] \leq \rho^+[X] < \infty$. If $G^-_X(z) < 0$, the condition $G^-_X(z) < \frac{F_X(z)}{1-F_X(z)}$ becomes non-binding, resulting in $\rho^+[X] = -\infty$; thus it must hold $G^-_X(z) \ge 0$ for all $z \in \R$. Conversely, if $(1-F_X(z))G^-_X(z) \geq 1$, then $\{z \in \R : G^-_X(z) <\frac{F_X(z)}{1-F_X(z)}\} = \emptyset$, leading to $\rho^+[X] = \infty$; thus it must hold $G^-_X(z) < \frac{1}{1-F_X(z)}$ for all $z \in \R$. A similar reasoning applies to $G^+_X$. Consequently, the functions $G^-_X$ and $G^+_X$ are non-negative and  strictly bounded by $\frac{1}{1-F_X}$.
\hfill \Halmos
\endproof

Next we provide examples of risk measures that fall within our framework and construct novel elicitable risk measures. Within the examples we also provide the generalised quantile representation discussed in this section.

\section{Examples: old and new}\label{sec:examples}
We first explore known elicitable functionals, and show that the mean, quantiles, expectiles, shortfall risk measures, and generalised quantiles all fall within our framework. 

\begin{example}[mean]\label{ex:mean}
    The expectation falls within our framework with $\phi(x) = x^2$ and $\mu(z, x) = 1_{x \neq z}$, leading to the well-known representation
    \begin{equation*}
        \E[X] = \argmin_{x \in \R} \E[(z-X)^2]\,.
    \end{equation*}
\end{example}

\begin{example}[Quantiles]\label{ex.var}  
The quantile, also known as the Value-at-Risk (VaR) at level \(\alpha\), arises with \(\phi(x) = x\) and 
\begin{equation*}
\mu_{\text{VaR}_\alpha}(z, x) = 
\begin{cases}
\alpha & \text{if } x > z, \\
1 - \alpha & \text{if } x < z, \\
0 & \text{if } x = z.
\end{cases}
\end{equation*}
The set \(\rho[X] = \argmin_{z \in \mathbb{R}} \mathbb{E}\big[\alpha(X - z)^+ + (1 - \alpha)(X - z)^-\big]\) represents the set of \(\alpha\)-quantiles of \(X\). The left and right endpoint of $\rho$, i.e. \(\rho^-\) and \(\rho^+\), correspond to the left and right quantiles, respectively.  
Calculations show that $G_X^+(z) = \frac{\alpha}{1 - \alpha}$ and consequently, the VaR can be expressed as
\[
\text{VaR}_\alpha(X) = \rho^-[X] = \inf\Big\{z \in \mathbb{R}: \frac{F_X(z)}{1 - F_X(z)} \ge G_X^+(z)\Big\} 
=
\inf\big\{z \in \mathbb{R}: F_X(z) \ge \alpha\big\}\,.
\]

The score \(\S\) is continuous, convex, and positive homogeneous, since  \(\phi\) is positive homogeneous, and $\mu_{\VaR}$ satisfies 
\[
\mu_{\VaR}(z, x + c) = \mu_{\VaR}(z - c, x) \quad \text{and} \quad \mu_{\VaR}(\lambda z, \lambda x) = \lambda^0 \mu_{\VaR}(z, x).
\]
By \Cref{prop: basic properties of functional}, VaR is positive homogeneous, translation invariant, and monotone. However, the derivative of the function \(z \mapsto \mathbb{E}[\S(z, X)]\) is not strictly increasing. This aligns with the well-known fact that quantiles are not always unique, nor are they convex.
\end{example}

\begin{example}[Generalised quantiles]\label{ex.generalized-quanti}
\cite{bellini2014generalized} studies generalized quantiles of the form
\begin{equation}\label{eq:generalised-quantile}
\argmin_{z \in \mathbb{R}} \mathbb{E} \big[\alpha \, \phi_1 \big((X - z)^+\big) + (1 - \alpha) \, \phi_2 \big((X - z)^-\big) \big] \,,    
\end{equation}
where $(\cdot)^+ := \max\{\cdot, 0\}$, $(\cdot)^- := -\min\{\cdot, 0\}$ denotes the positive and negative part, 
and where \(\phi_1, \phi_2\) are convex, strictly increasing functions such that \(\phi_1(0) = \phi_2(0) = 0\) and \(\phi_1(1) = \phi_2(1) = 1\). Under \(\mu_{\text{VaR}_\alpha}\), our functional is 
\begin{equation}\label{eq:new-generalised-quantile}
    \rho[X] = \argmin_{z \in \mathbb{R}} \mathbb{E} \big[\alpha \, \phi \big((X - z)^+\big) + (1 - \alpha) \, \phi \big((X - z)^-\big) \big]\,,
\end{equation}
which is a generalized quantile in the sense of \cite{bellini2014generalized,breckling1988m} with \(\phi = \phi_1 = \phi_2\). Thus our functional encompasses the expectiles  with $\phi(x) = x^2$ \citep{newey1987asymmetric}, the $L^p$-quantiles with $\phi(x) = x^p$, for $p \geq 2$ \citep{chen1996conditional}, and any positive homogeneous generalized quantile.

Any scoring function constructed via \eqref{eq:new-generalised-quantile} has the same properties as VaR. If moreover \(\phi\) is differentiable then
\[
G_X^+(z) = \frac{\alpha \mathbb{E}[\phi'(X - z) \mid X \geq z]}{(1 - \alpha) \mathbb{E}[\phi'(z - X) \mid X < z]},
\]
and by \Cref{theo: inf rep},
\[
\rho^-[X] = \inf \Big\{z \in \mathbb{R} ~\big|~ F_X(z) \geq \dfrac{\alpha \mathbb{E}[\phi'(X - z) \mid X \geq z]}{\alpha \mathbb{E}[\phi'(X - z) \mid X \geq z] + (1 - \alpha) \mathbb{E}[\phi'(z - X) \mid X < z]} \Big\}.
\]

Our approach is, in fact, more general than the one in \cite{bellini2014generalized}. We can recover the case where \(\phi_1 \neq \phi_2\) by modifying the \(\mu\) function. For example, let \(\phi_1, \phi_2\) be as in their study, and let \(\phi(x) \geq \max\{\alpha \phi_1(x), (1 - \alpha) \phi_2(x)\}\) for all positive real numbers, with \(\phi\) increasing faster than both \(\phi_1\) and \(\phi_2\). Additionally, define
\[
\mu(z, x) = 
\begin{cases}
\dfrac{\alpha \phi_1(x - z)}{\phi(x - z)} & \text{if } x > z, \\
\dfrac{(1 - \alpha) \phi_2(z - x)}{\phi(z - x)} & \text{if } x < z, \\
0 & \text{if } x = z.
\end{cases}
\]
Then, it holds that
\[
\rho[X] = \argmin_{z \in \mathbb{R}} \mathbb{E} \big[\phi(|X - z|) \mu(z, \lambda X) \big] = \argmin_{z \in \mathbb{R}} \mathbb{E} \big[\alpha \phi_1 \big((X - z)^+\big) + (1 - \alpha) \phi_2 \big((X - z)^-\big) \big]\,,
\]
which is precisely the same functional as in \cite{bellini2014generalized}. In which case, if $\phi_1 $ and $\phi_2$ are differentiable, then
\[
\rho^-[X] = \inf \Big\{z \in \mathbb{R} : F_X(z) \geq \dfrac{\alpha \mathbb{E}[\phi_1'(X - z) \mid X \geq z]}{\alpha \mathbb{E}[\phi_1'(X - z) \mid X \geq z] + (1 - \alpha) \mathbb{E}[\phi_2'(z - X) \mid X < z]} \Big\}.
\]
\end{example}

\begin{example}[Lambda quantiles]\label{ex. lambda quant}
     The Lambda quantile is defined by
    \begin{align*}
        \Lambda VaR (X) = \inf\{z \in \R : F_X(z) > \Lambda(z)\},
    \end{align*}
    for a monotone and right-continuous function $\Lambda: \R \to [ \lambda_M,\lambda_m]$, $0<\lambda_m<\lambda_M<1$. Lambda quantiles have  been studied in \cite{frittelli2014risk,burzoni2017properties,bellini2022axiomatization} and \cite{Ince2022QF}. The Lambda quantile is elicitable \citep{burzoni2017properties} and falls within our framework with $\phi(x) = x$ and 
    \begin{align*}
        \mu(z,x) = 
        \begin{cases}
\dfrac{\int_{z}^{x}\Lambda (y) \, dy}{\phi(x-z)}\quad \quad & \text{if}\quad  x > z\,, \\ 
1-\dfrac{\int_{x}^{z}\Lambda (y) \, dy}{\phi(z-x)}& \text{if} \quad x < z\,,
\\
0& \text{if}\quad x=z\,.
\end{cases}
    \end{align*}
Indeed, by \Cref{theo: inf rep}, it holds that $G_X^-(z) =\frac{\Lambda(z)}{1-\Lambda(z)}  $, which implies that
\begin{align*}
      \rho^+[X] 
      =
      \inf\big\{z \in \R: \frac{F_X(z)}{1-F_X(z)}> G_X^-(z) \big\} 
      = \inf\big\{z \in \R: {F_X(z)}> {\Lambda(z)} \big\} 
      =  \Lambda VaR (X)\,.
\end{align*}
In general, the $\Lambda \VaR$ is not a generalized quantile as defined in Bellini, i.e. it does not have representation \eqref{eq:generalised-quantile}. Its score \(\S\) is lower-semicontinuous, convex, and $\mu$ satisfies $
\mu(z, x + c) = \mu(z - c, x)$. By \Cref{prop: basic properties of functional}, $\Lambda$VaR is monotone, however, $\Lambda$VaR is positive homogeneous only if $\Lambda$ is constant, and under additional conditions on $\Lambda$, it is translation invariant.
\end{example}

\begin{example}[Shortfall]
    Shortfall risk measures were introduced by \cite{follmer2002convex}, and take the form 
    \begin{equation*}
        \rho[X] = \inf \{ z \in \R : \E[l(X-z)] \leq 0\}\,,
    \end{equation*}
    where $l : \R \rightarrow \R$ is increasing and non-constant, and we can assume without loss of generality that $l(-z) < 0 < l(z)$ for all $z >0$. \cite{bellini2015elicitable} showed that, as long as $l$ is left continuous and strictly increasing on $(-\infty, \ep)$ or on $(-\ep, \infty)$ for some $\ep>0$, then $\rho$ is elicitable. Now, take a continuous gauge function $\psi : \R \rightarrow [1,\infty)$ such that $ \frac{ -\int^0_{-x} l(y)dy}{ \int^x_0 l(y)dy} \leq \psi(x) $ for all $x >0$, then the shortfall risk measures are within our framework with $\phi (x) = \psi(x) \int_0^x l(y)dy$ and 
    \begin{align*}
        \mu(z,x) = \begin{cases}
            \dfrac{\int_0^{x-z} l(y)dy}{\psi(z-x)\int_0^{z-x} l(y)dy} &\text{ if } x<z \\ 
            \dfrac{1}{\psi(x-z)} &\text{ if } x>z \\
            0 &\text{ if } x=z .
        \end{cases}
    \end{align*}
The gauge function guarantees that $\mu \leq 1$. The well-known entropic risk measure is a shortfall risk measure with $l(z)=e^{\beta z}-1$, in this case we can take $\psi$ identically equal to $1$. In fact, $\frac{ -\int^0_{-x} l(y)dy}{ \int^x_0 l(y)dy} \leq \frac{-l(-x)}{l(x)}$, which is bounded by $1$ whenever $l$ is convex in which case $\psi \equiv 1$. As $l$ is monotone, it follows that $\S$ is convex and  \Cref{prop: existence} holds.
\end{example}

Our proposed risk measure thus covers the majority of elicitable risk measure in the literature. Indeed as elicitable coherent risk measures are a subclass of expectiles \citep{Steinwart2014proceeding}, and elicitable convex risk measures is a sub-family of shortfall risk measures \citep{bellini2015elicitable}, our new risk measures cover all convex and coherent elicitable risk measures.

Our framework allows to easily construct new elicitable functionals. Indeed any choice of combinations of $\phi$ and $\mu$, gives raise to an elicitable risk measure via \Cref{eq:rm-def}. Furthermore, if specific properties of the elicitable risk measures are desirable, then by \Cref{prop: basic properties}, $\phi$ and $\mu$ can be chosen accordingly. Next, we provide a new class of elicitable risk measures.

\begin{example}
    Take $\phi$ as in \Cref{ex.generalized-quanti} and $\mu$ as in \Cref{ex. lambda quant}. The elicitable functional then takes the form
\begin{align*}
     \rho[X] = \argmin_{z \in \R}\; \E \Bigg[ \,\dfrac{\int_{z}^{X}\Lambda (y) dy}{\phi(X-z)} \phi \big((X-z)^+\big)  +\Big(1-\dfrac{\int_{X}^{z}\Lambda (y) dy}{\phi(z-X)}\Big) \phi \big((X-z)^-\big) \,\Bigg]\,,
\end{align*}
Additionally, $G^-_X (z) = \frac{\Lambda (z)}{ \E [ \phi^\prime (z-X)| X<z] - \Lambda (z)}$ and \Cref{theo: inf rep} gives $\rho^+(X) = \inf \big\{ z \in \R : F_X(z) > \frac{\Lambda(z)}{\E [ \phi^\prime (z-X)| X<z]}\big\}.$
The elicitable risk measure has the same properties as the one in \Cref{ex. lambda quant}.
\end{example}

\section{Extensions}\label{sec:extensions}

In this section we explore two direction of extending our framework. First, using transformation of random variables, we obtain that any moment falls within our framework and provide a different representation of the entropic risk measure. Second, we illustrate how our elicitable risk measure can be extended to higher-order elicitability, that is how to, e.g., jointly elicit different quantile levels or the (mean, variance) pair. Moreover, this methodology can be used to jointly elicit any of the proposed new risk measures.

\subsection{Transformations}
Some functionals can be elicited by transforming the random variable of an elicitable risk measures in the following way.

\begin{lemma}[Transforming random variables]\label{lemma:trans-rv}
Let $\rho[\cdot]$ be elicitable with scoring function $\S(\cdot, \cdot)$. Then for a function $\ell \colon \R \to \R$, it holds that
\begin{equation*}
    \rho\big[\ell(X)\big] = \argmin_{z \in \R}\E\big[\S_\ell(z, X)\big]\,,
\end{equation*}
where $\S_\ell(z,x):= \S\big(z, \ell(x)\big) = \phi\big(\big|z - \ell(x)\big|\big)\, \mu(z, \ell(x))$. 
\end{lemma}
\proof{Proof.} 
    This follows immediately by the definition of elicitability, i.e.,
    \begin{align*}
        \rho\big[\ell(X)\big] = \argmin_{z \in \R}\E\big[\S\big(z, \ell(X)\big) \big]
        =
        \argmin_{z \in \R}\E\big[\S_\ell(z, X) \big]\,.
    \end{align*}%
    \hfill \Halmos
\endproof
The score $\S_\ell$ may preserve some of the properties of $\S$. For example, if $\ell$ is affine, then convexity of $\S$ implies convexity of $\S_\ell$. Furthermore if $\ell$ is affine and $\mu $ satisfies $\mu(z,x+c)=\mu(z-c,x)$, then $\mu(z,\ell(x+c)) = \mu(z,\ell(x)+l(c))=\mu(z-l(c),\ell(x))$. Thus, in this case, $\rho$ is translation invariant and $\rho[\ell(X + c)] = \rho[\ell(X)] + \ell(c)$. Additionally, $\S_\ell$ is positive homogeneity as long as $\S$ and $\ell$ are positive homogeneous.

A classical example is the $k$-th moment, $k \in \N$. Applying \Cref{lemma:trans-rv}, it holds that
\begin{equation}\label{eq:elicit-k-moment}
    \E[X^k] 
    =
    \argmin_{x \in \R}\E\big[\big(z - X^k)^2\big]
    = \argmin_{x \in \R}\E\big[\S_\ell(z, X)\big]
    \,,
\end{equation}
where $\S_\ell(z, x):= \S(z, \ell(x))$, with $\ell(x) = x^k$, $\phi(x) = x^2$, and $\mu(z,x) = 1_{z \neq x}$.

Another example includes the entropic risk measure, i.e. $\rho^\gamma[X] = \frac{1}{\gamma} \log\big(\E[e^{\gamma X}]\big)$, for $\gamma >0$. Indeed
\begin{equation*}
    \rho^\gamma[X]
    =
    \tfrac{1}{\gamma} \log\big(\E[e^{\gamma X}]\big)
    =\argmin_{z \in \R}\; \E\big[\S^{}_{e^\gamma \cdot}(z, X)\big]
    =\argmin_{z \in \R}\; \E\big[\S(e^{\gamma z}, e^{\gamma X})\big]
    \,,
\end{equation*}
where $\S_{e^{\gamma \cdot}}(z, x):= \S(e^{\gamma z}, e^{\gamma x})$, with $\phi(x) = x^2$, and $\mu(z,x) = 1_{z \neq x}$. This follows by applying \Cref{lemma:trans-rv} with $\ell(x) = e^{\gamma x}$, and \Cref{lemma-Osband}, discussed in the next section, with $g(x) = \frac{1}{\gamma} \log(x)$.

\subsection{Higher-order elicitability}
In this section, we establish how the proposed elicitable functionals can be generalised to higher-order elicitability. We first recall the definition of higher-order elicitability for a general functional and refer the interested reader to \cite{Lambert2008ACM} and \cite{fissler2016AS}.

\begin{definition}[$k$-elicitability]
\label{def:elicitable-multi}
For an action domain $\bA \subseteq \R^k$, $k \in \N$, a score is a measurable map $\bsS: \bA\times \R \to [0,\infty]$. Given a law-invariant, potentially set-valued, functional $R: \X \to \bA $, a score is called 
\begin{enumerate}[label = \roman*)]
    \item consistent for $R$, if for all $X\in \X$ and for all $\z:=(z_1, \ldots, z_k) \in \bA $ it holds that
    \begin{equation}\label{eq:consistency-higher-order}
        \E\big[ \bsS\big(R[X],X\big) \big]  \leq  \E \big[ \bsS(\z,X) \big] \,.
    \end{equation}

    \item strictly consistent for $R$, if equality in \eqref{eq:consistency-higher-order} holds only if $\z \in  R[X]$.
\end{enumerate}
We say that $R$ is $k$-elicitable, if there exists a consistent scoring function $ \bsS \colon \bA \times \R \to [0, \infty]$ for $R$, in which case $R$ admits representation
\begin{equation*}
    R[X]  = \argmin_{\z \in \bA} \E \big[ \bsS(\z,X) \big] \quad \text{for all} \quad X \in \X\,.
\end{equation*}
\end{definition}

The subclass of higher-order elicitable risk functionals that are composed of $1$-elicitable risk functionals, falls within our framework.

For this, let $\rho_j$, $j = 1, \ldots, k$, be elicitable functionals with corresponding scoring function $\S_j(z, x) := \phi_j(|x - z|)\mu_j(z, x)$. Then, by lemma 2.6 in \cite{fissler2016AS} it holds that
\begin{equation}\label{eq:score-additive}
    \big(\rho_1[X], \ldots, \rho_k[X]\big)
    =
    \argmin_{\z \in \R^k} \; \E\Big[\sum_{j = 1}^k \S_j(z_j, X)\Big]\,.
\end{equation}
In particular, the functional $\big(\rho_1, \ldots, \rho_k\big)$ is $k$-elicitable with the additive scoring function $\bS(\z, y):= \sum_{j = 1}^k \S_j(z_j, X)$, where $\rho_j$ can be any of the new elicitable functionals given in \eqref{eq:rm-def}.

Under mild conditions, \cite{Lambert2008ACM} show in theorem 5, that a scoring function $\bsS\colon \bA \to \R$, that elicits a $k$-dimensional functional, is accuracy rewarding for $(R_1, \ldots, R_k)$ if and only if $\bsS(\z, x) = \sum_{j = 1}^k S_j (z, x)$, where for each $j \in \{1, \ldots, k\}$, $S_j$ is a consistent scoring function for $R_j$. This shows that additive scores, and thus the functionals defined in \eqref{eq:score-additive} are the only $k$-elicitable functionals that lie within our setting.

\begin{example}[First and second moment]\label{ex:1-2moment}
    A straightforward example of a separable scoring function in \eqref{eq:score-additive} is the joint elicitability of the first and second moment. Recall that 
    \begin{equation*}
        \E[X] = \argmin_{z \in \R} \E[\S_1(z, X)]
        \quad \text{and} \quad
        \E[X^2] = \argmin_{z \in \R} \E[\S_2(z, X)]\,,
    \end{equation*}
    where $\S_1(z, x) = (|z-x|)^2$, with $\phi_1(x) = x^2$ and $\mu_1(z,x) = 1_{z \neq x}$, and where $\S_2(z,x) = \S_{(\cdot)^2}(z,x) = (|z-x^2|)^2$ with $\phi_2(\cdot) = \phi_1(\cdot)$ and $\mu_1(\cdot, \cdot) =\mu_2(\cdot, \cdot)   $, see \Cref{ex:mean} and \Cref{eq:elicit-k-moment}. Thus, the first two moments are jointly elicitable as, see also \cite{Gneiting2011},
    \begin{equation}\label{eq:1-2-moment-joint}
        \big(\E[X] ,\, \E[X^2]\big)
        =
        \argmin_{\z \in \R^2}\; \E\big[\S_1(z_1, X) + \S_2(z_2, X)\big]\,.
    \end{equation}
\end{example}

Next, we use Osband's principle to show that, e.g, the pair (mean, variance)  falls into our setting. For this we first recall
Osband's principle \citep{Lambert2008ACM,Gneiting2011}. 

\begin{lemma}[Osband's principle]\label{lemma-Osband}
    Let $(\rho_1, \ldots, \rho_k)$ be $k$-elicitable with scoring function $\bsS(\z, x)$ (not necessarily separable) and $g \colon \R^k \to \R^k$ be a bijective function. Then
    \begin{equation*}
        g\big((\rho_1[X], \ldots, \rho_k[X])\big)
        =
        \argmin_{z \in \R^k} \; \E\big[\bsS^g(\z, X) \big]\,,
    \end{equation*}
    where $\bsS^g(\z, x):= \bsS(g^{-1}(\z), x) $. 
\end{lemma}

\begin{example}[Mean and variance]
Continuing from \Cref{ex:1-2moment}, we apply the bijective map $g(z_1, z_2) = (z_1,\, z_2 - z_1^2)$ to \eqref{eq:1-2-moment-joint} and use \Cref{lemma-Osband} to obtain that
\begin{align*}
    \big(\E[X], \, \var(X)\big)
    =
    \argmin_{\z \in \R^2}\; \E\big[\S_1(z_1, X) + \S_2(z_2 + z_1^2, X)\big]
\end{align*}
where $\var(X):= \E[(X - \E[X])^2]$.
\end{example}

Similar approaches could be used to jointly elicit different quantile levels or jointly elicit any of the newly proposed elicitable risk measures.

\section{Conclusion}\label{sec:conclusion}
In this work, we propose a framework to explicitly construct old and new elicitable risk measures with specific properties such as monotonicity, translation invariance, and / or convexity, using multiplicative scoring functions. We show that the elicitable functionals are well-defined and prove necessary conditions on the component of the scoring function so that the functional has the above properties. Thus, this work is a recipe for constructing new elicitable functionals in an explicit fashion, and opens doors for a multitude of applications ranging from financial risk management to statistics and machine learning.

%
%
%


\ACKNOWLEDGMENT{SP would like acknowledge support from the Natural Sciences and Engineering Research Council of Canada with grant numbers DGECR-2020-00333 and RGPIN-2020-04289.}


\bibliographystyle{informs2014} 
\bibliography{refs} 

\begin{thebibliography}{28}
\providecommand{\natexlab}[1]{#1}
\providecommand{\url}[1]{\texttt{#1}}
\providecommand{\urlprefix}{URL }

\bibitem[{Acerbi \protect\BIBand{} Szekely(2014)}]{Acerbi2014Risk}
Acerbi C, Szekely B (2014) Back-testing {E}xpected {S}hortfall. \emph{Risk} 27(11):76--81.

\bibitem[{Artzner et~al.(1999)Artzner, Delbaen, Eber, \protect\BIBand{} Heath}]{ArtznerDelbaenETAL1999}
Artzner P, Delbaen F, Eber JM, Heath D (1999) Coherent measures of risk. \emph{Mathematical Finance} 9:203--228.

\bibitem[{Bellini \protect\BIBand{} Bignozzi(2015)}]{bellini2015elicitable}
Bellini F, Bignozzi V (2015) On elicitable risk measures. \emph{Quantitative Finance} 15(5):725--733.

\bibitem[{Bellini et~al.(2014)Bellini, Klar, M{\"u}ller, \protect\BIBand{} Gianin}]{bellini2014generalized}
Bellini F, Klar B, M{\"u}ller A, Gianin ER (2014) Generalized quantiles as risk measures. \emph{Insurance: Mathematics and Economics} 54:41--48.

\bibitem[{Bellini \protect\BIBand{} Peri(2022)}]{bellini2022axiomatization}
Bellini F, Peri I (2022) An axiomatization of $\lambda$-quantiles. \emph{SIAM Journal on Financial Mathematics} 13(1):SC26--SC38.

\bibitem[{Breckling \protect\BIBand{} Chambers(1988)}]{breckling1988m}
Breckling J, Chambers R (1988) M-quantiles. \emph{Biometrika} 75(4):761--771.

\bibitem[{Burzoni et~al.(2017)Burzoni, Peri, \protect\BIBand{} Ruffo}]{burzoni2017properties}
Burzoni M, Peri I, Ruffo C (2017) On the properties of the {L}ambda {V}alue at {R}isk: robustness, elicitability and consistency. \emph{Quantitative Finance} 17(11):1735--1743.

\bibitem[{Chen(1996)}]{chen1996conditional}
Chen Z (1996) Conditional {Lp}-quantiles and their application to the testing of symmetry in non-parametric regression. \emph{Statistics \& {P}robability letters} 29(2):107--115.

\bibitem[{Fissler \protect\BIBand{} Pesenti(2023)}]{fissler2023EJOR}
Fissler T, Pesenti SM (2023) Sensitivity measures based on scoring functions. \emph{European Journal of Operational Research} 307(3):1408--1423.

\bibitem[{Fissler \protect\BIBand{} Ziegel(2016)}]{fissler2016AS}
Fissler T, Ziegel JF (2016) Higher order elicitability and {O}sband’s principle. \emph{The Annals of Statistics} 44(4):1680--1707.

\bibitem[{F{\"o}llmer \protect\BIBand{} Schied(2002)}]{follmer2002convex}
F{\"o}llmer H, Schied A (2002) Convex measures of risk and trading constraints. \emph{Finance and {S}tochastics} 6:429--447.

\bibitem[{Frittelli \protect\BIBand{} Gianin(2002)}]{Frittelli2002JBF}
Frittelli M, Gianin ER (2002) Putting order in risk measures. \emph{Journal of Banking \& Finance} 26(7):1473--1486.

\bibitem[{Frittelli et~al.(2014)Frittelli, Maggis, \protect\BIBand{} Peri}]{frittelli2014risk}
Frittelli M, Maggis M, Peri I (2014) Risk measures on $ \mathcal{P}(\mathds{R})$ and {V}alue at {R}isk with probability/loss function. \emph{Mathematical Finance} 24(3):442--463.

\bibitem[{Gneiting(2011)}]{Gneiting2011}
Gneiting T (2011) {Making and Evaluating Point Forecasts}. \emph{Journal of the American Statistical Association} 106(494):746--762.

\bibitem[{Harjulehto \protect\BIBand{} H{\"a}st{\"o}(2019)}]{harjulehto2019generalized}
Harjulehto P, H{\"a}st{\"o} P (2019) \emph{Generalized Orlicz Spaces} (Springer).

\bibitem[{Ince et~al.(2022)Ince, Peri, \protect\BIBand{} Pesenti}]{Ince2022QF}
Ince A, Peri I, Pesenti S (2022) Risk contributions of {L}ambda quantiles. \emph{Quantitative Finance} 22(10):1871--1891.

\bibitem[{Kannappan(2009)}]{kannappan2009functional}
Kannappan P (2009) \emph{Functional equations and inequalities with applications} (Springer Science \& Business Media).

\bibitem[{Kou \protect\BIBand{} Peng(2016)}]{kou2016OR}
Kou S, Peng X (2016) On the measurement of economic tail risk. \emph{Operations Research} 64(5):1056--1072.

\bibitem[{Lambert et~al.(2008)Lambert, Pennock, \protect\BIBand{} Shoham}]{Lambert2008ACM}
Lambert NS, Pennock DM, Shoham Y (2008) Eliciting properties of probability distributions. \emph{Proceedings of the 9th ACM Conference on Electronic Commerce}, 129--138.

\bibitem[{Meng et~al.(2023)Meng, Taylor, Ben~Taieb, \protect\BIBand{} Li}]{meng2023scores}
Meng X, Taylor JW, Ben~Taieb S, Li S (2023) Scores for multivariate distributions and level sets. \emph{Operations Research} .

\bibitem[{Newey \protect\BIBand{} Powell(1987)}]{newey1987asymmetric}
Newey WK, Powell JL (1987) Asymmetric least squares estimation and testing. \emph{Econometrica: Journal of the Econometric Society} 819--847.

\bibitem[{Neyman \protect\BIBand{} Roughgarden(2023)}]{neyman2023OR}
Neyman E, Roughgarden T (2023) From proper scoring rules to max-min optimal forecast aggregation. \emph{Operations Research} 71(6):2175--2195.

\bibitem[{Nolde \protect\BIBand{} Ziegel(2017)}]{Nolde2017AAS}
Nolde N, Ziegel JF (2017) Elicitability and backtesting: Perspectives for banking regulation. \emph{Annals of Applied Statistics} 11:1833--74.

\bibitem[{Pesenti et~al.(2024)Pesenti, Jaimungal, Saporito, \protect\BIBand{} Targino}]{pesenti2024OR}
Pesenti SM, Jaimungal S, Saporito YF, Targino RS (2024) Risk budgeting allocation for dynamic risk measures. \emph{Operations Research} Forthcoming.

\bibitem[{Pesenti \protect\BIBand{} Vanduffel(2024)}]{Pesenti2024ORL}
Pesenti SM, Vanduffel S (2024) Optimal transport divergences induced by scoring functions. \emph{Operations Research Letters} 57:107146.

\bibitem[{Smith \protect\BIBand{} Bickel(2022)}]{smith2022OR}
Smith ZJ, Bickel JE (2022) Weighted scoring rules and convex risk measures. \emph{Operations Research} 70(6):3371--3385.

\bibitem[{Steinwart et~al.(2014)Steinwart, Pasin, Williamson, \protect\BIBand{} Zhang}]{Steinwart2014proceeding}
Steinwart I, Pasin C, Williamson R, Zhang S (2014) Elicitation and identification of properties. \emph{Conference on Learning Theory}, 482--526 (PMLR).

\bibitem[{Ziegel(2016)}]{Ziegel2016}
Ziegel JF (2016) Coherence and elicitability. \emph{Mathematical Finance} 26(4):901--918.

\end{thebibliography}



\end{document}